\documentclass[submission, copyright, creativecommons]{eptcs}
\providecommand{\event}{QPL 2023} % Name of the event you are submitting to

\usepackage{iftex}
\usepackage[english]{babel}

\ifpdf
  \usepackage{underscore}         % Only needed if you use pdflatex.
  \usepackage[T1]{fontenc}        % Recommended with pdflatex
\else
  \usepackage{breakurl}           % Not needed if you use pdflatex only.
\fi

\usepackage{hyperref}
\usepackage{amsmath}
\usepackage{amsthm}
\usepackage{subcaption}
\usepackage[ruled]{algorithm2e}
\usepackage{multirow}
\usepackage[]{nicematrix}
\usepackage{makecell}
\usepackage{graphicx}

\usepackage[draft]{tikzit}
\tikzstyle{gate}=[shape=rectangle, text height=1.5ex, text depth=0.25ex, yshift=0.5mm, fill=white, draw=black, minimum height=5mm, yshift=-0.5mm, minimum width=5mm, font={\small}, tikzit category=circuit]
\tikzstyle{big gate}=[shape=rectangle, text height=1.5ex, text depth=0.25ex, yshift=0.5mm, fill=white, draw=black, minimum height=10mm, yshift=-0.5mm, minimum width=5mm, font={\small}, tikzit category=circuit]
\tikzstyle{Z dot}=[inner sep=0mm, minimum size=2mm, shape=circle, draw=black, fill=zxgreen, tikzit fill={rgb,255: red,221; green,255; blue,221}, tikzit category=zx]
\tikzstyle{Z bold dot}=[inner sep=0mm, minimum size=2mm, shape=circle, draw=black, fill=zxgreen, tikzit fill={rgb,255: red,221; green,255; blue,221}, line width=1.2pt, tikzit category=zx]
\tikzstyle{Z phase dot}=[minimum size=5mm, font={\footnotesize\boldmath}, shape=rectangle, rounded corners=2mm, inner sep=0.2mm, outer sep=-2mm, scale=0.8, tikzit shape=circle, draw=black, fill=zxgreen, tikzit fill={rgb,255: red,221; green,255; blue,221}, tikzit draw=blue, tikzit category=zx]
\tikzstyle{Z tiny dot}=[inner sep=0mm, minimum size=1mm, shape=circle, draw=black, fill=zxgreen, tikzit fill={rgb,255: red,221; green,255; blue,221}]
\tikzstyle{X dot}=[Z dot, shape=circle, draw=black, fill=zxred, tikzit fill={rgb,255: red,255; green,136; blue,136}, tikzit category=zx]
\tikzstyle{X bold dot}=[inner sep=0mm, minimum size=2mm, shape=circle, draw=black, fill=zxred, tikzit fill={rgb,255: red,255; green,136; blue,136}, line width=1.2pt, tikzit category=zx]
\tikzstyle{X phase dot}=[Z phase dot, tikzit shape=circle, tikzit draw=blue, fill=zxred, tikzit fill={rgb,255: red,255; green,136; blue,136}, font={\footnotesize\boldmath}, tikzit category=zx]
\tikzstyle{X tiny dot}=[inner sep=0mm, minimum size=1mm, shape=circle, draw=black, fill=zxred, tikzit fill={rgb,255: red,255; green,136; blue,136}]
\tikzstyle{hadamard}=[fill=yellow, draw=black, shape=rectangle, inner sep=0.6mm, minimum height=1.5mm, minimum width=1.5mm, tikzit category=zx]
\tikzstyle{paulibox}=[fill={rgb,255: red,221; green,221; blue,255}, draw=black, shape=rectangle, inner sep=0.6mm, minimum height=5mm, minimum width=5mm, font={\footnotesize}, text height=1.5ex, text depth=0.25ex, tikzit category=zx]
\tikzstyle{vertex}=[inner sep=0.2mm, minimum size=1mm, shape=circle, draw=black, fill=black, tikzit category=misc]
\tikzstyle{vertex set}=[inner sep=0.2mm, minimum size=1mm, shape=circle, draw=black, fill=white, font={\footnotesize\boldmath}, tikzit category=misc]
\tikzstyle{small black dot}=[fill=black, draw=black, shape=circle, inner sep=0pt, minimum width=1.2mm, tikzit category=circuit]
\tikzstyle{cnot ctrl}=[fill=black, draw=black, shape=circle, inner sep=0pt, minimum width=1.2mm, tikzit category=circuit]
\tikzstyle{cnot targ}=[fill=white, draw=white, shape=circle, tikzit category=circuit, label={center:$\oplus$}, inner sep=0pt, minimum width=2.1mm, tikzit fill={rgb,255: red,102; green,204; blue,255}, tikzit draw=black]
\tikzstyle{ket}=[fill=white, draw=black, shape=regular polygon, regular polygon sides=3, regular polygon rotate=-30, scale=0.7, inner sep=1pt, tikzit category=circuit, tikzit shape=rectangle, tikzit fill=green]
\tikzstyle{bra}=[fill=white, draw=black, shape=regular polygon, regular polygon sides=3, regular polygon rotate=30, scale=0.7, inner sep=1pt, tikzit category=circuit, tikzit shape=rectangle, tikzit fill=red]
\tikzstyle{scalar}=[shape=rectangle, text height=1.5ex, text depth=0.25ex, yshift=0.5mm, fill=white, draw=black, minimum height=5mm, yshift=-0.5mm, minimum width=5mm, font={\small}]
\tikzstyle{clabel}=[fill=white, draw=none, shape=rectangle, tikzit fill={rgb,255: red,56; green,255; blue,242}, font={\footnotesize}, inner sep=1pt, tikzit category=labels]
\tikzstyle{empty diagram}=[draw={gray!40!white}, dashed, shape=rectangle, minimum width=1cm, minimum height=1cm, tikzit category=misc]
\tikzstyle{amap}=[fill=white, draw=black, shape=NEbox, tikzit category=asymmetric, tikzit fill=yellow, tikzit shape=rectangle]
\tikzstyle{amap conj}=[fill=white, draw=black, shape=NWbox, tikzit category=asymmetric, tikzit fill=green, tikzit shape=rectangle]
\tikzstyle{amap adj}=[fill=white, draw=black, shape=SEbox, tikzit category=asymmetric, tikzit fill=red, tikzit shape=rectangle]
\tikzstyle{amap trans}=[fill=white, draw=black, shape=SWbox, tikzit category=asymmetric, tikzit fill=orange, tikzit shape=rectangle]
\tikzstyle{astate}=[fill=white, draw=black, shape=NEtriangle, tikzit category=asymmetric, tikzit shape=circle, tikzit fill=yellow]
\tikzstyle{astate conj}=[fill=white, draw=black, shape=NWtriangle, tikzit category=asymmetric, tikzit shape=circle, tikzit fill=green]
\tikzstyle{astate adj}=[fill=white, draw=black, shape=SEtriangle, tikzit category=asymmetric, tikzit shape=circle, tikzit fill=red]
\tikzstyle{astate trans}=[fill=white, draw=black, shape=SWtriangle, tikzit category=asymmetric, tikzit shape=circle, tikzit fill=orange]
\tikzstyle{box}=[shape=rectangle, text height=1.5ex, text depth=0.25ex, yshift=0.5mm, fill=white, draw=black, minimum height=5mm, yshift=-0.5mm, minimum width=5mm, font={\small}]
\tikzstyle{medium box}=[shape=rectangle, text height=1.5ex, text depth=0.25ex, yshift=0.5mm, fill=white, draw=black, minimum height=10mm, yshift=-0.5mm, minimum width=5mm, font={\small}]

\tikzstyle{simple}=[-]
\tikzstyle{hadamard edge}=[-, dashed, dash pattern=on 2pt off 0.5pt, thick, draw={rgb,255: red,68; green,136; blue,255}]
\tikzstyle{box edge}=[-, dashed, dash pattern=on 2pt off 0.5pt, thick, draw={rgb,255: red,203; green,192; blue,225}]
\tikzstyle{brace edge}=[-, tikzit draw=blue, decorate, decoration={brace,amplitude=1mm,raise=-1mm}]
\tikzstyle{diredge}=[->]
\tikzstyle{double edge}=[-, double, shorten <=-1mm, shorten >=-1mm, double distance=2pt]
\tikzstyle{gray edge}=[-, {gray!60!white}]
\tikzstyle{pointer edge}=[->, very thick, gray]
\tikzstyle{boldedge}=[-, line width=1.2pt, shorten <=-0.17mm, shorten >=-0.17mm]
\tikzstyle{bidir edge}=[<->, very thick, draw={rgb,255: red,191; green,191; blue,191}]
\tikzstyle{surface X}=[-, tikzit fill=red, fill=zxred]
\tikzstyle{surface Z}=[-, tikzit fill=green, fill=zxgreen]

\input{quantum.tikzdefs}

\theoremstyle{definition}
\newtheorem{theorem}{Theorem}[section]

\newtheorem{definition}[theorem]{Definition}

\newtheorem{example*}[theorem]{Example*}
\newtheorem{examples*}[theorem]{Examples*}

\newtheorem{remark*}[theorem]{Remark*}

\renewcommand{\>}{\rangle}

\newcommand{\rowop}[2]{\tikz{%
\node at (0,0.5) {\scriptsize $R_{#2}\! := \!R_{#1} \!+\! R_{#2}$};
\draw [-latex] (-0.75,0) -- (0.75,0);
}\xspace}

\newcommand{\matrixop}[1]{\tikz{%
\node at (0,0.5) {#1};
\draw [-latex] (-0.75,0) -- (0.75,0);
}\xspace}

\newcommand{\submatrixAi}{
  \begin{pNiceMatrix}[r,margin,first-row,first-col]
  \CodeBefore
  \cellcolor{gray!50}{2-,-6}% Colour the row and col that have been eliminated
  \Body
    & 0 & 1 & 2 & 3 & 4 & 5 & 6 & 7 \\
  0 & 0 & 0 & 0 & 1 & 1 & 0 & 1 & 1 \\
  5 & 0 & 0 & 0 & 0 & 0 & 1 & 0 & 0 \\
  7 & 1 & 1 & 0 & 1 & 1 & 0 & 1 & 1 \\
  3 & 1 & 1 & 0 & 1 & 0 & 0 & 0 & 0 \\
  \end{pNiceMatrix}
}

\newcommand{\submatrixAii}{
  \begin{pNiceMatrix}[r,margin,first-row,first-col]
  \CodeBefore
  \cellcolor{red!50}{2-,-6}% Colour the row and col that have been eliminated
  \Body
    & 0 & 1 & 2 & 3 & 4 & 5 & 6 & 7 \\
  0 & 0 & 0 & 0 & 1 & 1 & 0 & 1 & 1 \\
  5 & 0 & 0 & 0 & 0 & 0 & 1 & 0 & 0 \\
  7 & 1 & 1 & 0 & 1 & 1 & 0 & 1 & 1 \\
  3 & 1 & 1 & 0 & 1 & 0 & 0 & 0 & 0 \\
  \end{pNiceMatrix}
}

\newcommand{\submatrixBi}{
  \begin{pNiceMatrix}[r,margin,first-row,first-col]
  \CodeBefore
  \cellcolor{gray!50}{3-,-8}% Colour the row and col that have been eliminated
  \Body
    & 0 & 1 & 2 & 3 & 4 & 5 & 6 & 7 \\
  0 & 0 & 0 & 0 & 1 & 1 & 0 & 1 & 1 \\
  4 & 1 & 1 & 0 & 0 & 1 & 0 & 1 & 0 \\
  7 & 1 & 1 & 0 & 1 & 1 & 0 & 1 & 1 \\
  3 & 1 & 1 & 0 & 1 & 0 & 0 & 0 & 0 \\
  \end{pNiceMatrix}
}
\newcommand{\submatrixBii}{
  \begin{pNiceMatrix}[r,margin,first-row,first-col]
  \CodeBefore
  \cellcolor{red!50}{3-,-8}% Colour the row and col that have been eliminated
  \Body
    & 0 & 1 & 2 & 3 & 4 & 5 & 6 & 7 \\
  0 & 1 & 1 & 0 & 0 & 0 & 0 & 0 & 0 \\
  4 & 1 & 1 & 0 & 0 & 1 & 0 & 1 & 0 \\
  7 & 0 & 0 & 0 & 0 & 0 & 0 & 0 & 1 \\
  3 & 1 & 1 & 0 & 1 & 0 & 0 & 0 & 0 \\
  \end{pNiceMatrix}
}

\newcommand{\submatrixCi}{
  \begin{pNiceMatrix}[r,margin,first-row,first-col]
  \CodeBefore
  \cellcolor{gray!50}{3-,-7}% Colour the row and col that have been eliminated
  \Body
    & 0 & 1 & 2 & 3 & 4 & 5 & 6 & 7 \\
  0 & 1 & 1 & 0 & 0 & 0 & 0 & 0 & 0 \\
  4 & 1 & 1 & 0 & 0 & 1 & 0 & 1 & 0 \\
  6 & 1 & 1 & 0 & 1 & 0 & 0 & 1 & 0 \\
  3 & 1 & 1 & 0 & 1 & 0 & 0 & 0 & 0 \\
  \end{pNiceMatrix}
}
\newcommand{\submatrixCii}{
  \begin{pNiceMatrix}[r,margin,first-row,first-col]
  \CodeBefore
  \cellcolor{red!50}{3-,-7}% Colour the row and col that have been eliminated
  \Body
    & 0 & 1 & 2 & 3 & 4 & 5 & 6 & 7 \\
  0 & 1 & 1 & 0 & 0 & 0 & 0 & 0 & 0 \\
  4 & 0 & 0 & 0 & 1 & 1 & 0 & 0 & 0 \\
  6 & 0 & 0 & 0 & 0 & 0 & 0 & 1 & 0 \\
  3 & 1 & 1 & 0 & 1 & 0 & 0 & 0 & 0 \\
  \end{pNiceMatrix}
}

\newcommand{\submatrixDi}{
  \begin{pNiceMatrix}[r,margin,first-row,first-col]
  \CodeBefore
  \cellcolor{gray!50}{4-,-4}% Colour the row and col that have been eliminated
  \Body
    & 0 & 1 & 2 & 3 & 4 & 5 & 6 & 7 \\
  0 & 1 & 1 & 0 & 0 & 0 & 0 & 0 & 0 \\
  4 & 0 & 0 & 0 & 1 & 1 & 0 & 0 & 0 \\
  2 & 1 & 0 & 1 & 0 & 0 & 0 & 0 & 0 \\
  3 & 1 & 1 & 0 & 1 & 0 & 0 & 0 & 0 \\
  \end{pNiceMatrix}
}
\newcommand{\submatrixDii}{
  \begin{pNiceMatrix}[r,margin,first-row,first-col]
  \CodeBefore
  \cellcolor{red!50}{4-,-4}% Colour the row and col that have been eliminated
  \Body
    & 0 & 1 & 2 & 3 & 4 & 5 & 6 & 7 \\
  0 & 1 & 1 & 0 & 0 & 0 & 0 & 0 & 0 \\
  4 & 1 & 1 & 0 & 0 & 1 & 0 & 0 & 0 \\
  2 & 1 & 0 & 1 & 0 & 0 & 0 & 0 & 0 \\
  3 & 0 & 0 & 0 & 1 & 0 & 0 & 0 & 0 \\
  \end{pNiceMatrix}
}

\newcommand{\submatrixEi}{
  \begin{pNiceMatrix}[r,margin,first-row,first-col]
  \CodeBefore
  \cellcolor{gray!50}{2-,-5}% Colour the row and col that have been eliminated
  \Body
    & 0 & 1 & 2 & 3 & 4 & 5 & 6 & 7 \\
  0 & 1 & 1 & 0 & 0 & 0 & 0 & 0 & 0 \\
  4 & 1 & 1 & 0 & 0 & 1 & 0 & 0 & 0 \\
  2 & 1 & 0 & 1 & 0 & 0 & 0 & 0 & 0 \\
  3 & 0 & 0 & 0 & 1 & 0 & 0 & 0 & 0 \\
  \end{pNiceMatrix}
}
\newcommand{\submatrixEii}{
  \begin{pNiceMatrix}[r,margin,first-row,first-col]
  \CodeBefore
  \cellcolor{red!50}{2-,-5}% Colour the row and col that have been eliminated
  \Body
    & 0 & 1 & 2 & 3 & 4 & 5 & 6 & 7 \\
  0 & 1 & 1 & 0 & 0 & 0 & 0 & 0 & 0 \\
  4 & 0 & 0 & 0 & 0 & 1 & 0 & 0 & 0 \\
  2 & 1 & 0 & 1 & 0 & 0 & 0 & 0 & 0 \\
  3 & 0 & 0 & 0 & 1 & 0 & 0 & 0 & 0 \\
  \end{pNiceMatrix}
}

\newcommand{\submatrixFi}{
  \begin{pNiceMatrix}[r,margin,first-row,first-col]
  \CodeBefore
  \cellcolor{gray!50}{1-,-1}% Colour the row and col that have been eliminated
  \Body
    & 0 & 1 & 2 & 3 & 4 & 5 & 6 & 7 \\
  0 & 1 & 1 & 0 & 0 & 0 & 0 & 0 & 0 \\
  1 & 0 & 1 & 0 & 0 & 0 & 0 & 0 & 0 \\
  2 & 1 & 0 & 1 & 0 & 0 & 0 & 0 & 0 \\
  3 & 0 & 0 & 0 & 1 & 0 & 0 & 0 & 0 \\
  \end{pNiceMatrix}
}
\newcommand{\submatrixFii}{
  \begin{pNiceMatrix}[r,margin,first-row,first-col]
  \CodeBefore
  \cellcolor{red!50}{1-,-1}% Colour the row and col that have been eliminated
  \Body
    & 0 & 1 & 2 & 3 & 4 & 5 & 6 & 7 \\
  0 & 1 & 0 & 0 & 0 & 0 & 0 & 0 & 0 \\
  1 & 0 & 1 & 0 & 0 & 0 & 0 & 0 & 0 \\
  2 & 0 & 1 & 1 & 0 & 0 & 0 & 0 & 0 \\
  3 & 0 & 0 & 0 & 1 & 0 & 0 & 0 & 0 \\
  \end{pNiceMatrix}
}

\newcommand{\submatrixGi}{
  \begin{pNiceMatrix}[r,margin,first-row,first-col]
  \CodeBefore
  \cellcolor{gray!50}{2-,-2}% Colour the row and col that have been eliminated
  \Body
    & 0 & 1 & 2 & 3 & 4 & 5 & 6 & 7 \\
  0 & 1 & 0 & 0 & 0 & 0 & 0 & 0 & 0 \\
  1 & 0 & 1 & 0 & 0 & 0 & 0 & 0 & 0 \\
  2 & 0 & 1 & 1 & 0 & 0 & 0 & 0 & 0 \\
  3 & 0 & 0 & 0 & 1 & 0 & 0 & 0 & 0 \\
  \end{pNiceMatrix}
}
\newcommand{\submatrixGii}{
  \begin{pNiceMatrix}[r,margin,first-row,first-col]
  \CodeBefore
  \cellcolor{red!50}{2-,-2}% Colour the row and col that have been eliminated
  \Body
    & 0 & 1 & 2 & 3 & 4 & 5 & 6 & 7 \\
  0 & 1 & 0 & 0 & 0 & 0 & 0 & 0 & 0 \\
  1 & 0 & 1 & 0 & 0 & 0 & 0 & 0 & 0 \\
  2 & 0 & 0 & 1 & 0 & 0 & 0 & 0 & 0 \\
  3 & 0 & 0 & 0 & 1 & 0 & 0 & 0 & 0 \\
  \end{pNiceMatrix}
}

\title{Global Synthesis of CNOT Circuits with Holes}
\author{Ewan Murphy
\institute{University of Oxford}
\email{ewan.murphy360@gmail.com}
\and
Aleks Kissinger
\institute{University of Oxford}
\email{aleks.kissinger@cs.ox.ac.uk}
}
\def\titlerunning{Global Synthesis of CNOT Circuits with Holes}
\def\authorrunning{E. Murphy \& A. Kissinger}

\date{}

\begin{document}

\maketitle

\begin{abstract}
  % Most quantum computing algorithms of interest were developed with fault-tolerant computers in mind. Faced with having only Noisy Intermediate Scale Quantum (NISQ) computers to run these algorithms, a variety of methods have been developed to transform quantum circuits into forms better suited to the hardware accessible to us.

  A common approach to quantum circuit transformation is to use the properties of a specific gate set to create an efficient representation of a given circuit's unitary, such as a parity matrix or stabiliser tableau, and then resynthesise an improved circuit, e.g. with fewer gates or respecting connectivity constraints. Since these methods rely on a restricted gate set, generalisation to arbitrary circuits usually involves slicing the circuit into pieces that can be resynthesised and working with these separately. The choices made about what gates should go into each slice can have a major effect on the performance of the resynthesis. In this paper we propose an alternative approach to generalising these resynthesis algorithms to general quantum circuits. Instead of cutting the circuit into slices, we ``cut out'' the gates we can't resynthesise leaving holes in our quantum circuit. The result is a second-order process called a quantum comb, which can be resynthesised directly. We apply this idea to the RowCol algorithm, which resynthesises CNOT circuits for topologically constrained hardware, explaining how we were able to extend it to work for quantum combs. We then compare the generalisation of RowCol using our method to the na\"ive ``slice and build'' method empirically on a variety of circuit sizes and hardware topologies. Finally, we outline how quantum combs could be used to help generalise other resynthesis algorithms.
\end{abstract}

\section{Introduction}

Current quantum computers suffer from severe limitations such as high error rates, low numbers of qubits, and connectivity constraints for multi-qubit operations. Furthermore, without error correction, the poor fidelity of current gate implementations compounds over the execution of the circuit, so it advantageous to find the smallest possible circuit to represent an algorithm.

Using the properties of a specific gate set, an efficient representation of a circuit's unitary can be resynthesised into an improved circuit. For example, the unitary action of CNOT circuits can be fully described by the associated $\mathbb F_2$-linear function over basis vectors on $n$-qubit space, seen as vectors in $\mathbb F_2^n$. In other words, we can represent the action of a CNOT circuit using a matrix over $\mathbb F_2$, called its \textit{parity matrix}.

Noting that the parity matrix of a single CNOT gate corresponds to an elementary row operation, it is possible to resynthesise a CNOT circuit from its parity matrix by performing Gauss-Jordan elimination to reduce the matrix to identity and introducing one CNOT gate for each corresponding row operation. This method was introduced in~\cite{alber2001quantum} and refined in the Patel-Markov-Hayes algorithm~\cite{PatelCNOT}, which produces asymptotically optimal gate counts for (unconstrained) CNOT synthesis. Similar ideas involving the decomposition of symplectic matrices into basic generators have also been applied for synthesising Clifford circuits from stabiliser tableaux~\cite{aaronson2004improved,maslov2018shorter,dehaene2003clifford,vandennest2010classical}.

Some quantum computers, such as some superconducting devices~\cite{stassi2020scalable}, have the additional limitation of restricted connectivity, meaning 2-qubit gates aren't allowed between arbitrary pairs of qubits. By restricting which row operations are possible when reducing a parity matrix to the identity, circuits that obey specific connectivity constraints can be synthesised from ones that don't. A variety of techniques for introducing these constraints based on Steiner trees~\cite{nash2020quantum,kissinger2019cnot,rowcol,permrowcol} and integer programming~\cite{de2022decoding} have been introduced in recent years.

\begin{figure}[ht]
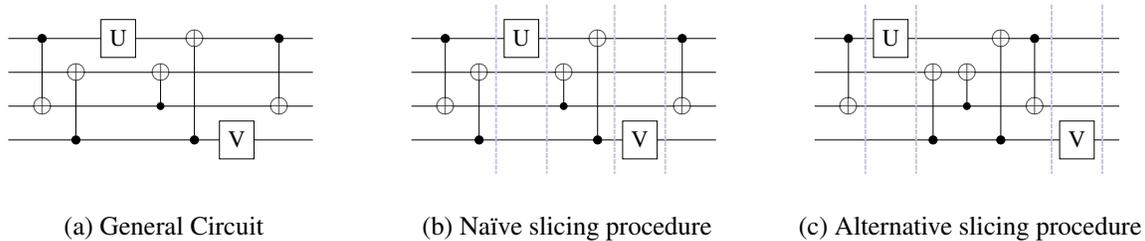

  \begin{subfigure}{0.33\textwidth}
    \ctikzfig{slicingexampleA}
    \caption{General Circuit}
    \label{slicecircA}
  \end{subfigure}
  \begin{subfigure}{0.33\textwidth}
    \ctikzfig{slicingexampleB}
    \caption{Na\"ive slicing procedure}
    \label{slicecircB}
  \end{subfigure}
  \begin{subfigure}{0.33\textwidth}
    \ctikzfig{slicingexampleC}
    \caption{Alternative slicing procedure}
    \label{slicecircC}
  \end{subfigure}
  \caption{Possible slicing procedures for a synthesis process that can't deal with hadamard gates}
  \label{slicecircuits}
\end{figure}

These synthesis methods provide a way to optimise circuits that mitigates some of the current limitations of NISQ devices. However, they suffer from relying on the properties of a specific gate set. The conventional generalisation of these methods to arbitrary quantum circuits is to slice the circuit into pieces that can be synthesised, and treat each of them separately\cite{gheorghiu2022reducing}. This isn't as straightforward as it might initially sound. Consider for example the circuit in Figure~\Ref{slicecircA} and a CNOT circuit synthesis procedure. The synthesis procedure can't deal with the circuit as it currently is due to the gates U and V, so we need to slice the circuit into sections containing just CNOT gates and synthesise those. As some of the gates in a quantum circuit can be moved past each other, there isn't always a unique way to make a slice. Take the two possible slicing options in Figures~\Ref{slicecircB} and \Ref{slicecircC} for example. Since the CNOT slices are treated independently during the synthesis procedure, which sides of the slice certain CNOTs end up on can affect the size of the new circuit by allowing/preventing simplifications or cancellations.

In this paper we present a new approach to generalising circuit synthesis that doesn't rely on slicing the circuit into pieces. Instead we remove the gates that can't be synthesised, leaving holes in the circuit and producing what is known as a quantum comb~\cite{chiribella2008quantum,eggeling2002semicausal}. This quantum comb can be understood as a new circuit with additional qubits, where these new qubits represent the old qubits at different points in time. We then explain how extending the functionality of the CNOT synthesis algorithm RowCol to work quantum combs allows one to route general circuits. Our generalisation is then empirically tested against the na\"ive slicing process on a variety of circuit sizes and hardware constraints, finding our method has increasingly better performance than the slicing one as circuit sizes increase. Finally, we outline how quantum combs could be used to generalise other circuit synthesis procedures, as well as ways in which our current algorithm could be further optimised.

This paper has the following structure. Section~\Ref{sec:paritymatrix} explains how the parity matrix representation of a CNOT circuit works. Section~\Ref{sec:rowcol} introduces the idea behind CNOT synthesis algorithms, with a focus on the circuit routing algorithm RowCol. In Section~\Ref{sec:combs} we introduce quantum combs, as well as language specific to this paper that will be helpful in explaining our algorithm. Section~\Ref{sec:combsynth} explains our extension to RowCol that allows working with quantum combs. Empirical results comparing our quantum combs method to the slicing process are presented in Section~\Ref{sec:resandcon}. Finally, Section~\Ref{sec:futurework} concludes the paper and outlines possible future works for using quantum combs to generalise other synthesis procedures.

\section{Preliminaries}
\subsection{CNOT Circuit as a Parity Matrix}\label{sec:paritymatrix}
In this paper, we will use the phrases ``CNOT circuit'' or ``CNOT comb'' which refer to circuits or quantum combs entirely made of CNOTS. CNOT stands for ``controlled not'' and is a quantum gate that acts on 2-qubits: the control $c$ and the $t$, CNOT($c$, $t$). It acts in such a way that a NOT gate is applied to the target qubit only if the control qubit is in state $|1\>$, CNOT$|0\>|0\> = |0\>|0\>$ and CNOT$|1\>|0\> = |1\>|1\>$. When we refer to states in this section we mean computational basis states and the behaviour for general superpositions can be inferred from the linearity of quantum operations. An alternative way of thinking about about CNOTs is as modulo 2 addition: the target qubit changes state to the sum of the control and target values modulo 2, CNOT$|c\>|t\> = |c\>|c\oplus t\>$. This idea is represented as a circuit diagram in Figure~\Ref{fig:cnotmoduloexample}. A CNOT gate can therefore be written as a list of which qubits are present in the parity equations of the output states: a representation of this is given by the matrix over $\mathbb{F}_{2}$ in Figure~\Ref{fig:cnotparitymatrix}. By reasoning about CNOT gates in this way we can write an entire CNOT circuit as a list of parity equations, Figure~\Ref{fig:paritycircuitexample}, then represent that circuit by an invertible element of $\mathbb F^{n\times n}_{2}$ (i.e a parity matrix), Figure~\Ref{fig:paritymatrixexample}. One way to construct this matrix is by traversing the circuit and applying row operations for each CNOT, CNOT($c$, $t$) corresponds to R($c$, $t$), where R($c$, $t$) means setting row $t$ the sum of rows $c$ and $t$ modulo 2. By identifying a set of row operations that reduce a parity matrix to the identity, a CNOT circuit can be generated: this is the core principle behind the CNOT circuit synthesis algorithms discussed in the next section.

\begin{figure}[ht]
  \begin{subfigure}{0.5\textwidth}
    \ctikzfig{cnotparity}
    \caption{CNOT gate as modulo 2 addition of basis states}\label{fig:cnotmoduloexample}
  \end{subfigure}
  \begin{subfigure}{0.5\textwidth}
    \[
      \begin{pmatrix}
        1 & 0  \\
        1 & 1  \\
      \end{pmatrix}
    \]
    \caption{Parity matrix for a CNOT gate}\label{fig:cnotparitymatrix}
  \end{subfigure}
  \caption{Parity representations of a CNOT gate}\label{fig:cnotrep}
\end{figure}

\begin{figure}[ht]
  \begin{subfigure}{0.5\textwidth}
    \ctikzfig{paritycircuitexample}
    \caption{CNOT circuit described as parity equations}
    \label{fig:paritycircuitexample}
  \end{subfigure}
  \begin{subfigure}{0.5\textwidth}
    \[
      \mathbf{P}
      \ =\
      \begin{pmatrix}
        1 & 0 & 0 & 1 \\
        1 & 1 & 1 & 1 \\
        0 & 0 & 1 & 1 \\
        0 & 0 & 0 & 1 \\
      \end{pmatrix}
    \]
    \caption{Parity matrix corresponding to circuit in~\Ref{fig:paritycircuitexample}}
    \label{fig:paritymatrixexample}
  \end{subfigure}
  \caption{Parity representation of a CNOT circuit}\label{fig:cnotcircrep}
\end{figure}

%These equations can then be written as a matrix shown in Figure~\Ref{fig:paritymatrixexample}, the columns correspond to the output of the qubits and the rows to the inputs. Where there is a 1, the qubit for this row will be included in the equation for this column. Knowing this it can easily be seen how the equations in Figure~\Ref{fig:paritycircuitexample} become the matrix in Figure~\Ref{fig:paritymatrixexample}. Another way to construct $\mathbf{P}$ would be by starting with the identity and performing row operations for each CNOT in the circuit, CNOT(c,t) corresponds to the c-th row being added to the t-th row. This idea can be performed in reverse, determining which row operations convert a parity matrix to the identity generates its CNOT circuit.

\subsection{CNOT Circuit Synthesis Algorithms}\label{sec:rowcol}

Synthesis algorithms provide a means of reducing the size of CNOT circuits\cite{PatelCNOT}, or allow the resynthesis of circuits under topological constraints\cite{kissinger2019cnot,nash2020quantum,permrowcol}. The ability to perform both of these tasks efficiently is vital for NISQ computing, as it provides a way to best utilise the machines we have available whilst minimising the consequences of their limitations. These CNOT synthesis methods work by converting the circuit into a parity matrix, as described in Section~\ref{sec:paritymatrix}, then identifying which row operations convert the matrix back to the identity. This sequence of row operations corresponds to the generated CNOT circuit.

% \begin{figure}[ht]
%   \ctikzfig{rowcoltopology}
%   \caption{Example graph of which qubits CNOTs can be applied between}
%   \label{nisqtopology}
% \end{figure}

For some quantum computers, superconducting ones, for example \cite{stassi2020scalable}, CNOT gates may be restricted to only be possible between nearest neighbour qubits. These computers are said to be topologically constrained, with graphs representing the allowed CNOTs called the topology, and the problem of converting a circuit to one that obeys these constraints known as quantum circuit routing. The systematic introduction of SWAPs into the circuit is one way to overcome this problem\cite{cowtan2019qubit}. However, in this paper we will focus on the circuit synthesis approaches to finding a solution, specifically the ones that use parity matrices, as there are other alternative approaches that synthesise from different representations \cite{de2022decoding}. By restricting the possible row operations when reducing a parity matrix to the identity, circuit synthesis methods can be also used to produce circuits that obey topological constraints. One of these synthesis algorithms is known as RowCol, and will be the focus of the rest of this section.

Starting with a CNOT circuit $\mathcal C$, and a graph $G(V,E)$ representing connectivity of qubits, RowCol synthesises a new circuit as follows:

\begin{algorithm}[H]
\caption{RowCol}
\label{alg:rowcol}
\SetKwInOut{Input}{Input}
\SetKwInOut{Output}{Output}
\Input{A circuit $\mathcal C$, and topology $G(V,E)$}
\Output{A circuit $\mathcal C'$, respecting the topological constraints }
\begin{enumerate}
  \item Generate a new empty circuit $\mathcal C'$ with the same number of qubits as $\mathcal C$.
  \item Compute the parity matrix $P$ of $\mathcal C$.
  \item Pick a non-cutting vertex of $V$ of $G$, and get its corresponding qubit $q$.
  \item Apply elementary row operations, restricted to the edges of $G(V,E)$, to reduce row $q$ and \\ column $q$ to a unit vector. %Eliminate row $q$ and column $q$ of $P$ using Steiner trees.
  \item Remove vertex $V$ from $G$, and row and column $q$ from $P$.
  \item Go back to Step 2 and repeat until $G$ has no more vertices.
  \item Return $\mathcal C'$
\end{enumerate}
\end{algorithm}

Note the column of $q$ is reduced to a unit vector by using row operations to place a $1$ on the diagonal, then adding row $q$ to the other rows to eliminate 1s above and below the diagonal. The row can be made into a unit vector by solving a system of linear equations for other rows to add back on to row $q$.

In order to respect connectivity constraints, row operations might not be performed directly, but via some intermediate operations computed using Steiner trees. While this is an important aspect of RowCol and related algorithms, we can treat this process essentially as a ``black box'' for the purposes of our algorithm. We refer readers to the paper that introduced RowCol~\cite{rowcol} or other Steiner-tree based algorithms~\cite{kissinger2019cnot,nash2020quantum,permrowcol} for details.

We are now going to step through the RowCol procedure for the matrix in Figure~\ref{fig:paritymatrixexample}. We start with qubit 0, meaning we are going to need to eliminate the 0th column, then the 0th row. To eliminate the column we need to perform $R(0,1)$, and to eliminate the row we need to perform $R(3,0)$. This has then reduced the 0th column and row to unit vectors, therefore we can ignore these when eliminating the rest of the matrix.
\[
\begin{pNiceMatrix}[r,margin]
\CodeBefore
\cellcolor{gray!50}{2-1,3-1,4-1}
\Body
1 & 0 & 0 & 1 \\
1 & 1 & 1 & 1 \\
0 & 0 & 1 & 1 \\
0 & 0 & 0 & 1 \\
\end{pNiceMatrix}
\rowop{0}{1}
\begin{pNiceMatrix}[r,margin]
\CodeBefore
\cellcolor{gray!50}{1-2,1-3,1-4}
\Body
1 & 0 & 0 & 1 \\
0 & 1 & 1 & 0 \\
0 & 0 & 1 & 1 \\
0 & 0 & 0 & 1 \\
\end{pNiceMatrix}
\rowop{3}{0}
\begin{pNiceMatrix}[r,margin]
\CodeBefore
\cellcolor{red!50}{1-1,1-2,1-3,1-4,2-1,3-1,4-1}
\Body
1 & 0 & 0 & 0 \\
0 & 1 & 1 & 0 \\
0 & 0 & 1 & 1 \\
0 & 0 & 0 & 1 \\
\end{pNiceMatrix}
\]
The column for the 1st qubit is already a unit vector, meaning we can move directly onto the row. To eliminate this row we need to perform $R(2,1)$ and $R(3,1)$: although a specific order is shown on the diagram either will work to eliminate the row. As both the column and row are now eliminated, we can ignore these in the subsequent operations.
\[
\begin{pNiceMatrix}[r,margin]
\CodeBefore
\cellcolor{gray!50}{2-3,2-4}
\Body
1 & 0 & 0 & 0 \\
0 & 1 & 1 & 0 \\
0 & 0 & 1 & 1 \\
0 & 0 & 0 & 1 \\
\end{pNiceMatrix}
\rowop{2}{1}
\begin{pNiceMatrix}[r,margin]
\CodeBefore
\cellcolor{gray!50}{2-3,2-4}
\Body
1 & 0 & 0 & 0 \\
0 & 1 & 0 & 1 \\
0 & 0 & 1 & 1 \\
0 & 0 & 0 & 1 \\
\end{pNiceMatrix}
\rowop{3}{1}
\begin{pNiceMatrix}[r,margin]
\CodeBefore
\cellcolor{red!50}{2-2,2-3,2-4,3-2,4-2}
\Body
1 & 0 & 0 & 0 \\
0 & 1 & 0 & 0 \\
0 & 0 & 1 & 1 \\
0 & 0 & 0 & 1 \\
\end{pNiceMatrix}
\]
The column is already a unit vector again here, meaning we move to the row, which can be eliminated by $R(3,2)$. This reduces the matrix to the identity meaning the process is over.
\[
\begin{pNiceMatrix}[r,margin]
\CodeBefore
\cellcolor{gray!50}{3-4}
\Body
1 & 0 & 0 & 0 \\
0 & 1 & 0 & 0 \\
0 & 0 & 1 & 1 \\
0 & 0 & 0 & 1 \\
\end{pNiceMatrix}
\rowop{3}{2}
\begin{pNiceMatrix}[r,margin]
\CodeBefore
\cellcolor{red!50}{3-3,3-4,4-3}
\Body
1 & 0 & 0 & 0 \\
0 & 1 & 0 & 0 \\
0 & 0 & 1 & 0 \\
0 & 0 & 0 & 1 \\
\end{pNiceMatrix}
\]

\begin{figure}[ht]
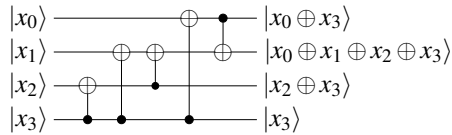

  \ctikzfig{rowcolgeneratedcircuit}
  \caption{Circuit generated from the step-by-step RowCol procedure}
  \label{fig:rowcolgeneratedcircuit}
\end{figure}

The sequence of row operations that converted the matrix to the identity are: $R(0,1)$, $R(3,0)$, $R(2,1)$, $R(3,1)$, $R(3,2)$. This tells us how to construct the resynthesised circuit which is shown in Figure~\ref{fig:rowcolgeneratedcircuit}.

\subsection{Quantum Combs}\label{sec:combs}

\begin{figure}
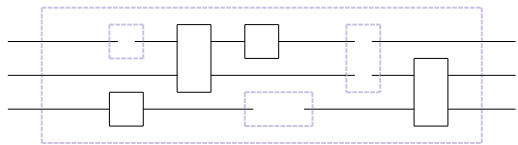

  \ctikzfig{generalQuantumComb}
  \caption{General representation of a quantum comb as quantum circuit with holes}\label{generalcomb}
\end{figure}

Here we will introduce the concept of a quantum comb, as well as the language and notation used in this paper to discuss specific aspects of them. A quantum comb is a generalisation of a quantum channel that can take other quantum channels as input, rather than states. These can be depicted graphically, as in Figure~\ref{generalcomb}, as circuits which not only have open wires at the top and bottom, but certain ``holes'' in the middle, where other gates can be inserted. The term ``comb'' comes from the fact that the entire object should no longer be represented as a box, but as an irregular shape resembling a hair comb, where each the of ``teeth'' corresponds to a distinct time step. See e.g.~\cite{chiribella2008quantum} for the formal definition and many such pictures.

Rather than defining combs in general, we will focus on combs arising from quantum circuits where certain single-qubit gates have been removed. The restriction to single-qubit holes is not essential, but it suffices for our purposes and will make certain aspects of our algorithm simpler. These can be described as normal quantum circuits with some extra information about time ordering of qubits, subject to some constraints. To motivate the definition, we will look at an example circuit, consisting of CNOT gates and several single-qubit gates we wish to remove:
\begin{equation}\label{quantumcircuithighlighted}
  \begin{tikzpicture}
	\begin{pgfonlayer}{nodelayer}
		\node [style=none] (0) at (-12, 2) {};
		\node [style=none] (1) at (-12, 0) {};
		\node [style=none] (2) at (-12, -2) {};
		\node [style=none] (3) at (-12, -4) {};
		\node [style=none] (4) at (18, 2) {};
		\node [style=none] (5) at (18, 0) {};
		\node [style=none] (6) at (18, -2) {};
		\node [style=none] (7) at (18, -4) {};
		\node [style=cnot ctrl] (8) at (-11, 2) {};
		\node [style=cnot targ] (9) at (-11, -2) {};
		\node [style=cnot targ] (10) at (-9, 2) {};
		\node [style=cnot ctrl] (11) at (-9, 0) {};
		\node [style=cnot ctrl] (12) at (-7, -4) {};
		\node [style=cnot targ] (13) at (-7, -2) {};
		\node [style=gate] (14) at (-9, -2) {U};
		\node [style=gate] (15) at (-7, 0) {V};
		\node [style=cnot ctrl] (16) at (-5, 2) {};
		\node [style=cnot targ] (17) at (-5, -2) {};
		\node [style=cnot targ] (18) at (-3, 0) {};
		\node [style=cnot ctrl] (19) at (-3, -2) {};
		\node [style=cnot ctrl] (20) at (-1, 2) {};
		\node [style=cnot targ] (21) at (-1, -4) {};
		\node [style=gate] (22) at (1, -2) {W};
		\node [style=cnot ctrl] (23) at (5, 0) {};
		\node [style=cnot targ] (24) at (5, -2) {};
		\node [style=cnot targ] (25) at (5, -2) {};
		\node [style=cnot ctrl] (26) at (7, 2) {};
		\node [style=cnot targ] (27) at (7, 0) {};
		\node [style=gate] (30) at (15, 0) {H};
		\node [style=cnot targ] (31) at (17, 2) {};
		\node [style=cnot ctrl] (32) at (17, -2) {};
		\node [style=none] (33) at (-7.75, 0.75) {};
		\node [style=none] (34) at (-6.25, 0.75) {};
		\node [style=none] (35) at (-7.75, -0.75) {};
		\node [style=none] (36) at (-6.25, -0.75) {};
		\node [style=none] (37) at (-9.75, -1.25) {};
		\node [style=none] (38) at (-8.25, -1.25) {};
		\node [style=none] (39) at (-9.75, -2.75) {};
		\node [style=none] (40) at (-8.25, -2.75) {};
		\node [style=none] (41) at (0.25, -1.25) {};
		\node [style=none] (42) at (1.75, -1.25) {};
		\node [style=none] (43) at (0.25, -2.75) {};
		\node [style=none] (44) at (1.75, -2.75) {};
		\node [style=none] (45) at (14.25, 0.75) {};
		\node [style=none] (46) at (15.75, 0.75) {};
		\node [style=none] (47) at (14.25, -0.75) {};
		\node [style=none] (48) at (15.75, -0.75) {};
		\node [style=none] (49) at (-13, 2) {0};
		\node [style=none] (50) at (-13, 0) {1};
		\node [style=none] (51) at (-13, -2) {2};
		\node [style=none] (52) at (-13, -4) {3};
		\node [style=none] (53) at (19, 2) {0};
		\node [style=none] (54) at (19, 0) {1};
		\node [style=none] (55) at (19, -2) {2};
		\node [style=none] (56) at (19, -4) {3};
		\node [style=cnot targ] (57) at (3, -2) {};
		\node [style=cnot ctrl] (58) at (3, -4) {};
		\node [style=cnot ctrl] (59) at (9, -2) {};
		\node [style=cnot targ] (60) at (9, 0) {};
		\node [style=cnot ctrl] (61) at (11, -4) {};
		\node [style=cnot targ] (62) at (11, -2) {};
		\node [style=cnot ctrl] (63) at (13, -2) {};
		\node [style=cnot targ] (64) at (13, 0) {};
	\end{pgfonlayer}
	\begin{pgfonlayer}{edgelayer}
		\draw (0.center) to (4.center);
		\draw (1.center) to (5.center);
		\draw (2.center) to (6.center);
		\draw (3.center) to (7.center);
		\draw (8) to (9);
		\draw (10) to (11);
		\draw (13) to (12);
		\draw (16) to (17);
		\draw (18) to (19);
		\draw (20) to (21);
		\draw (23) to (25);
		\draw (26) to (27);
		\draw (31) to (32);
		\draw [style=box edge] (33.center) to (34.center);
		\draw [style=box edge] (33.center) to (35.center);
		\draw [style=box edge] (35.center) to (36.center);
		\draw [style=box edge] (34.center) to (36.center);
		\draw [style=box edge] (37.center) to (38.center);
		\draw [style=box edge] (37.center) to (39.center);
		\draw [style=box edge] (39.center) to (40.center);
		\draw [style=box edge] (38.center) to (40.center);
		\draw [style=box edge] (41.center) to (42.center);
		\draw [style=box edge] (41.center) to (43.center);
		\draw [style=box edge] (43.center) to (44.center);
		\draw [style=box edge] (42.center) to (44.center);
		\draw [style=box edge] (45.center) to (46.center);
		\draw [style=box edge] (45.center) to (47.center);
		\draw [style=box edge] (47.center) to (48.center);
		\draw [style=box edge] (46.center) to (48.center);
		\draw (57) to (58);
		\draw (59) to (60);
		\draw (61) to (62);
		\draw (63) to (64);
	\end{pgfonlayer}
\end{tikzpicture}
\end{equation}
Note that we have labelled the inputs and outputs by the same indices $\{0,1,2,3\}$. We call these labels for the qubits in our original circuit the \textit{logical qubits}. If we remove the gates $U, V, W, H$ from the circuit, we obtain something that looks like this:
\begin{equation}\label{quantumcomb}
  \begin{tikzpicture}
	\begin{pgfonlayer}{nodelayer}
		\node [style=none] (0) at (-13, 2) {};
		\node [style=none] (1) at (-13, 0) {};
		\node [style=none] (2) at (-13, -2) {};
		\node [style=none] (3) at (-13, -4) {};
		\node [style=none] (4) at (17, 2) {};
		\node [style=none] (5) at (17, 0) {};
		\node [style=none] (6) at (17, -2) {};
		\node [style=none] (7) at (17, -4) {};
		\node [style=cnot ctrl] (8) at (-12, 2) {};
		\node [style=cnot targ] (9) at (-12, -2) {};
		\node [style=cnot targ] (10) at (-10, 2) {};
		\node [style=cnot ctrl] (11) at (-10, 0) {};
		\node [style=cnot ctrl] (12) at (-8, -4) {};
		\node [style=cnot targ] (13) at (-8, -2) {};
		\node [style=cnot ctrl] (16) at (-6, 2) {};
		\node [style=cnot targ] (17) at (-6, -2) {};
		\node [style=cnot targ] (18) at (-4, 0) {};
		\node [style=cnot ctrl] (19) at (-4, -2) {};
		\node [style=cnot ctrl] (20) at (-2, 2) {};
		\node [style=cnot targ] (21) at (-2, -4) {};
		\node [style=cnot ctrl] (23) at (4, 0) {};
		\node [style=cnot targ] (24) at (4, -2) {};
		\node [style=cnot targ] (25) at (4, -2) {};
		\node [style=cnot ctrl] (26) at (6, 2) {};
		\node [style=cnot targ] (27) at (6, 0) {};
		\node [style=cnot targ] (31) at (16, 2) {};
		\node [style=cnot ctrl] (32) at (16, -2) {};
		\node [style=none] (49) at (-14, 2) {0};
		\node [style=none] (50) at (-14, 0) {1};
		\node [style=none] (51) at (-14, -2) {2};
		\node [style=none] (52) at (-14, -4) {3};
		\node [style=none] (53) at (-9, 0) {};
		\node [style=none] (54) at (-7, 0) {};
		\node [style=none] (60) at (-11, -2) {};
		\node [style=none] (61) at (-9, -2) {};
		\node [style=none] (67) at (-1, -2) {};
		\node [style=none] (68) at (1, -2) {};
		\node [style=none] (74) at (13, 0) {};
		\node [style=none] (75) at (15, 0) {};
		\node [style=none] (81) at (-8.5, 0) {1};
		\node [style=none] (82) at (-7.5, 0) {4};
		\node [style=none] (83) at (-10.5, -2) {2};
		\node [style=none] (84) at (-9.5, -2) {6};
		\node [style=none] (85) at (13.5, 0) {4};
		\node [style=none] (86) at (14.5, 0) {5};
		\node [style=none] (87) at (18, 0) {5};
		\node [style=none] (88) at (18, -2) {7};
		\node [style=none] (89) at (18, -4) {3};
		\node [style=none] (90) at (18, 2) {0};
		\node [style=none] (91) at (-0.5, -2) {6};
		\node [style=none] (92) at (0.5, -2) {7};
		\node [style=cnot targ] (93) at (2, -2) {};
		\node [style=cnot ctrl] (94) at (2, -4) {};
		\node [style=cnot ctrl] (95) at (8, -2) {};
		\node [style=cnot targ] (96) at (8, 0) {};
		\node [style=cnot ctrl] (97) at (10, -4) {};
		\node [style=cnot targ] (98) at (10, -2) {};
		\node [style=cnot ctrl] (99) at (12, -2) {};
		\node [style=cnot targ] (100) at (12, 0) {};
	\end{pgfonlayer}
	\begin{pgfonlayer}{edgelayer}
		\draw (0.center) to (4.center);
		\draw (3.center) to (7.center);
		\draw (8) to (9);
		\draw (10) to (11);
		\draw (13) to (12);
		\draw (16) to (17);
		\draw (18) to (19);
		\draw (20) to (21);
		\draw (23) to (25);
		\draw (26) to (27);
		\draw (31) to (32);
		\draw (1.center) to (53.center);
		\draw (54.center) to (74.center);
		\draw (75.center) to (5.center);
		\draw (2.center) to (60.center);
		\draw (61.center) to (67.center);
		\draw (68.center) to (6.center);
		\draw (93) to (94);
		\draw (95) to (96);
		\draw (97) to (98);
		\draw (99) to (100);
	\end{pgfonlayer}
\end{tikzpicture}
\end{equation}
We can break the data represented by the picture above into two parts. First, we can see this as just a normal quantum circuit $\mathcal C$, but now acting on more qubits:
\begin{equation}\label{quantumcombunfolded}
  \begin{tikzpicture}
	\begin{pgfonlayer}{nodelayer}
		\node [style=none] (0) at (-15, 6) {};
		\node [style=none] (1) at (-15, 4) {};
		\node [style=none] (2) at (-15, 2) {};
		\node [style=none] (3) at (-15, 0) {};
		\node [style=none] (4) at (15, 6) {};
		\node [style=none] (5) at (15, -4) {};
		\node [style=none] (6) at (15, -8) {};
		\node [style=none] (7) at (15, 0) {};
		\node [style=cnot ctrl] (8) at (-14, 6) {};
		\node [style=cnot targ] (9) at (-14, 2) {};
		\node [style=cnot targ] (10) at (-12, 6) {};
		\node [style=cnot ctrl] (11) at (-12, 4) {};
		\node [style=cnot ctrl] (12) at (-10, 0) {};
		\node [style=cnot targ] (13) at (-10, -6) {};
		\node [style=cnot ctrl] (14) at (-8, 6) {};
		\node [style=cnot targ] (15) at (-8, -6) {};
		\node [style=cnot targ] (16) at (-6, -2) {};
		\node [style=cnot ctrl] (17) at (-6, -6) {};
		\node [style=cnot ctrl] (18) at (-4, 6) {};
		\node [style=cnot targ] (19) at (-4, 0) {};
		\node [style=cnot ctrl] (20) at (2, -2) {};
		\node [style=cnot targ] (21) at (2, -8) {};
		\node [style=cnot targ] (22) at (2, -8) {};
		\node [style=cnot ctrl] (23) at (4, 6) {};
		\node [style=cnot targ] (24) at (4, -2) {};
		\node [style=cnot targ] (25) at (14, 6) {};
		\node [style=cnot ctrl] (26) at (14, -8) {};
		\node [style=none] (27) at (-16, 6) {0};
		\node [style=none] (28) at (-16, 4) {1};
		\node [style=none] (29) at (-16, 2) {2};
		\node [style=none] (30) at (-16, 0) {3};
		\node [style=none] (31) at (15, 4) {};
		\node [style=none] (32) at (-15, -2) {};
		\node [style=none] (33) at (15, 2) {};
		\node [style=none] (34) at (-15, -6) {};
		\node [style=none] (35) at (15, -6) {};
		\node [style=none] (36) at (-15, -8) {};
		\node [style=none] (37) at (15, -2) {};
		\node [style=none] (38) at (-15, -4) {};
		\node [style=none] (39) at (16, 4) {1};
		\node [style=none] (40) at (-16, -2) {4};
		\node [style=none] (41) at (16, 2) {2};
		\node [style=none] (42) at (-16, -6) {6};
		\node [style=none] (43) at (16, -2) {4};
		\node [style=none] (44) at (-16, -4) {5};
		\node [style=none] (45) at (16, -4) {5};
		\node [style=none] (46) at (16, -8) {7};
		\node [style=none] (47) at (16, 0) {3};
		\node [style=none] (48) at (16, 6) {0};
		\node [style=none] (49) at (16, -6) {6};
		\node [style=none] (50) at (-16, -8) {7};
		\node [style=cnot targ] (51) at (0, -8) {};
		\node [style=cnot ctrl] (52) at (0, 0) {};
		\node [style=cnot ctrl] (53) at (6, -8) {};
		\node [style=cnot targ] (54) at (6, -2) {};
		\node [style=cnot targ] (55) at (8, -8) {};
		\node [style=cnot ctrl] (56) at (8, 0) {};
		\node [style=cnot ctrl] (57) at (10, -8) {};
		\node [style=cnot targ] (58) at (10, -2) {};
	\end{pgfonlayer}
	\begin{pgfonlayer}{edgelayer}
		\draw (0.center) to (4.center);
		\draw (3.center) to (7.center);
		\draw (8) to (9);
		\draw (10) to (11);
		\draw (13) to (12);
		\draw (14) to (15);
		\draw (16) to (17);
		\draw (18) to (19);
		\draw (20) to (22);
		\draw (23) to (24);
		\draw (25) to (26);
		\draw (1.center) to (31.center);
		\draw (32.center) to (37.center);
		\draw (38.center) to (5.center);
		\draw (2.center) to (33.center);
		\draw (34.center) to (35.center);
		\draw (36.center) to (6.center);
		\draw (51) to (52);
		\draw (53) to (54);
		\draw (55) to (56);
		\draw (57) to (58);
	\end{pgfonlayer}
\end{tikzpicture}
\end{equation}
Notice how this circuit is over more qubits than just the original logical qubits. The indices above refer to a qubit at a particular point in time, and hence have a many-to-one relationship with the logical qubits. We call these the \textit{temporal qubits}.

There is also a binary relation telling us which temporal qubits come directly before others, which we call the \textit{holes}. We can represent the holes indicated in \eqref{quantumcomb} as the set $\mathcal H = \{ (1,4), (2,6), (6,7), (4,5) \}$.

% \begin{definition}
%   A pair $(\mathcal C, \mathcal H)$ of a circuit on a set of qubits $Q = \{0,\ldots,n-1\}$ and a binary relation $\mathcal H \subseteq Q \times Q$ is called a \textit{comb} if there exists a partial ordering on $\mathcal H$
% \end{definition}

We can also define what it means to plug single-qubit gates into each of these holes. That is, for any set $\mathcal G$ of single-qubit gates and a \textit{plugging map} $p : \mathcal H \to \mathcal G$, we can unambiguously reconstruct a circuit. For example, we can reconstruct our original circuit using the plugging map
\[ p :: \{ (1,4) \mapsto V, (4,5) \mapsto H, (2,6) \mapsto U, (6,7) \mapsto W \}. \]

To define what it means to be a valid comb, we will formalise the process of plugging in gates, and require this to result in a well-defined circuit.

\begin{algorithm}[h]
\caption{Comb composition}
\label{alg:plugging}
\SetKwInOut{Input}{Input}
\SetKwInOut{Output}{Output}
\Input{A pair $(\mathcal C, \mathcal H)$ of a circuit $\mathcal{C}$ on a set of qubits $Q = \{0,\ldots,n-1\}$ and a binary relation $\mathcal H \subseteq Q \times Q$, as well as a plugging function $p: \mathcal H \to \mathcal G$}
\Output{A circuit $\mathcal C'$ or FAIL}

\bigskip

For each $(q_1, q_2) \in \mathcal H$:
\begin{enumerate}
  \item Re-order non-interacting gates in $\mathcal C$ such that all gates on qubit $q_1$ appear before gates on \\ qubit $q_2$. If this is not possible, FAIL.
  \item Insert $p((q_1, q_2))$ into $\mathcal C$ directly before the first gate to refer to qubit $q_2$.
  \item Remove $(q_1, q_2)$ from $\mathcal H$.
  \item Rename $q_2 \mapsto q_1$ in $\mathcal C$ and $\mathcal H$ and remove $q_2$ from $\mathcal C$.
  \item If this renaming produces a hole of the form $(q,q) \in \mathcal H$, then FAIL, otherwise continue to \\ the next hole.
\end{enumerate}
Return $\mathcal C' := \mathcal C$.

\bigskip

\end{algorithm}

\begin{definition}\label{def:comb}
  A pair $(\mathcal C, \mathcal H)$ is called a \textit{comb} if Algorithm \ref{alg:plugging} succeeds for any plugging map $p$.
\end{definition}

Note that this Algorithm \ref{alg:plugging} can fail if the pair $(\mathcal C, \mathcal H)$ introduces cyclic dependencies between temporal qubits. However, when we obtain such pairs by ``cutting out'' the non-CNOT gates from a circuit, this algorithm will always succeed. We can formalise this ``cutting out'' process with the following procedure:

\begin{algorithm}[h]
\caption{Comb decomposition}
\label{alg:decomp}
\SetKwInOut{Input}{Input}
\SetKwInOut{Output}{Output}
\Input{A circuit $\mathcal C'$}
\Output{A comb $(\mathcal C, \mathcal H)$ and a plugging map $p: \mathcal H \to \mathcal G$.}

\begin{enumerate}
\item Create an empty comb $(\mathcal{C}, \mathcal H)$ where $\mathcal C$ is an empty circuit with as many qubits as $\mathcal C'$ and $\mathcal H = \{ \}$.
\item Create a mapping $t$ for temporal qubits, where initially $t(q) = q$ for all qubits $q \in \mathcal C'$ and \\ $p$ is the empty map $\mathcal H \to \mathcal G$.
  \item For each gate $g$ in $C'$:
        \begin{itemize}
          \item if $g$ is a CNOT on qubits $q_{1}$ and $q_{2}$, add $g$ to $\mathcal{C}$ on qubits $t(q_{1}), t(q_{2})$
          \item if $g$ is not a CNOT and acting on qubit $q$, introduce a fresh qubit $q'$ to $\mathcal{C}$, add $(t(q),q')$ to $\mathcal H$, set $p((t(q),q')) := g$, and let $t(q) := q'$
        \end{itemize}
  \item Return $(\mathcal{C}, \mathcal H)$ and $p$.
\end{enumerate}

\end{algorithm}

\section{Algorithm: CombSynth}\label{sec:combsynth}

Our main algorithm proceeds by decomposing a circuit into a comb consisting of just CNOT gates and a plugging map for all the additional single-qubit gates. It then resynthesises the comb using a variation of the RowCol algorithm which preserves the comb structure, then plugs the single-qubit gates back in to give a fully routed circuit.

% Here we introduce two algorithms CombSynth and CombRowCol, CombSynth is the higher level algorithm that maps unrouted circuits to routed ones by decomposing them into combs for the routing procedure. CombRowCol is algorithm that we have developed that allows us to route a quantum comb.
% \begin{algorithm}[H]
%   \SetKwInOut{Input}{Input}
%   \SetKwInOut{Output}{Output}
%   \SetKwFunction{decompose}{CombDecomposition}
%   \SetKwFunction{recompose}{CombRecomposition}
%   \SetKwFunction{combsynth}{CombSynth}
%   \caption{CombSynth\label{combsynth}}
%   \Input{Circuit $C$ and a topology $G(V_G,E)$}
%   \Output{Circuit $C'$}
%   \tcc{Decompose the circuit C into a comb and a list of hole plugs}
%   $\mathcal{C}, H_{p} \leftarrow$ \decompose{C}\;
%   \tcc{Route the comb to the provided topology}
%   $\mathcal{C}' \leftarrow$ \combsynth{$\mathcal{C}$, $G(V_{G},E)$}\;
%   \tcc{Recompose a new routed circuit from the routed comb and the hole plugs}
%   $C' \leftarrow$ \recompose{$\mathcal{C}'$, $H_{p}$}\;
%   \Return $C'$
% \end{algorithm}
%

% Once we have generated the circuits quantum comb, its needs resynthesising, for this we use a modified version of RowCol, as introduced in Section~\Ref{sec:rowcol}.

The RowCol algorithm works by eliminating qubits one by one. To apply this idea to quantum combs we need to know what it means to remove one temporal qubit at a time. As temporal qubits represent sections of the ``lifetime'' of the original logical qubits, not all of them can exist at the same time. This means that, although we may write the comb as one large circuit, it is not possible to perform CNOTs between all the temporal qubits at any given time. This idea needs to be carried forward into the synthesis by restricting row operations to only be between qubits in the same sections of time.

For a comb $(\mathcal C, \mathcal H)$, a temporal qubit $q$ is called \textit{available} if it does not appear as the first part of a hole. That is, there exists no $q'$ such that $(q, q') \in \mathcal H$. It is \textit{extractible} if its row and column can be made into a unit vector using the RowCol algorithm restricted only to row operations between available qubits.

Finally, to track topological constraints, which may be relevant for RowCol, we maintain a connection between logical and available temporal qubits. For each logical qubit $q_0$, let $t(q_0)$ be its associated available temporal qubit. That is, let $t(q_0) = q_0$ if $q_0$ does not appear in any holes, otherwise, let it be $q_k$ for the maximal transitive chain of holes $(q_0, q_1), (q_1, q_2), \ldots, (q_{k-1}, q_k)$. Therefore, the mapping $t$ is determined by a collection of holes, meaning as we update $\mathcal H$ in the algorithm below we are also updating $t$.

Our main algorithm, CombSynth, works as follows:

\begin{algorithm}[H]
\caption{CombSynth}
\label{alg:combsynth}
\SetKwInOut{Input}{Input}
\SetKwInOut{Output}{Output}
\Input{A comb $(\mathcal C, \mathcal H)$ with temporal qubits $Q = \{0, \ldots, n-1\}$, topology graph $G(V, E)$}
\Output{A comb $(\mathcal C', \mathcal H')$, respecting topological constraints for any plugging $p$}

\begin{enumerate}
\item Create a new comb $(\mathcal{C}', \mathcal H')$ with $\mathcal C'$ initially empty and $\mathcal H' = \mathcal H$
\item Compute the parity matrix $P$ of $\mathcal C$.
\item Identify an extractible temporal qubit $e$ in comb $(\mathcal C, \mathcal H)$.
\item Produce a rectangular sub-matrix $P'$ with columns the same as $P$ and rows labelled by $t(q)$ for each logical qubit $q$.
\item Run an iteration of RowCol on row $t(e)$ and column $e$ of $P'$, with topology $G$, updating $\mathcal C'$.
\item Update the corresponding rows of $P$ using $P'$.
% \item if $(e', e)$ in $\mathcal H$ then set $t(e) = e'$
\item Remove $e$ from the qubits of $\mathcal C$ and any hole of the form $(e',e)$ for some $e'$ from $\mathcal H$.
\item Repeat from Step 3 until no qubits remain in $\mathcal C$.
\item Return $(\mathcal C', \mathcal H')$.
\end{enumerate}

\end{algorithm}

Note that if $(\mathcal C, \mathcal H)$ arose from a circuit by cutting single-qubit gates out, as in Algorithm~\ref{alg:decomp}, there will always be at least one extractible temporal qubit. We simply need to choose one corresponding to the latest gate that has been cut out of the circuit.

The core of the RowCol algorithm is used to reduce the row and column of the chosen temporal qubit to unit vectors. Applying this procedure iteratively, removing the temporal qubits in an allowed order, and updating the sub-matrix as you go, will reduce the overall parity matrix to the identity whilst ensuring that no row operations happen between qubits that exist at different points in time. Therefore we have a procedure for synthesising quantum combs from their parity matrices. This method eliminates rows and columns of a matrix in the same way as RowCol, and the rows of rectangular sub-matrix $P'$ always correspond to the same logical qubits, even though we are swapping the temporal qubits in and out. Hence, our algorithm can be used to route to constrained topologies in the same way as RowCol, but now it can be done for general unitaries by converting to a quantum comb. Comparisons of our generalisation of RowCol to a simple slicing procedure are presented in the next section.

\begin{figure}[ht]
\[
\begin{pNiceMatrix}[r,margin,first-row,first-col]
\CodeBefore
\cellcolor{blue!50}{1-,6-,8-,4-}% Colour the rows that will go into the sub matrix
\Body
  & 0 & 1 & 2 & 3 & 4 & 5 & 6 & 7 \\
0 & 0 & 0 & 0 & 1 & 1 & 0 & 1 & 1 \\
1 & 0 & 1 & 0 & 0 & 0 & 0 & 0 & 0 \\
2 & 1 & 0 & 1 & 0 & 0 & 0 & 0 & 0 \\
3 & 1 & 1 & 0 & 1 & 0 & 0 & 0 & 0 \\
4 & 1 & 1 & 0 & 0 & 1 & 0 & 1 & 0 \\
5 & 0 & 0 & 0 & 0 & 0 & 1 & 0 & 0 \\
6 & 1 & 1 & 0 & 1 & 0 & 0 & 1 & 0 \\
7 & 1 & 1 & 0 & 1 & 1 & 0 & 1 & 1 \\
\end{pNiceMatrix}
\matrixop{Generate sub-matrix}
\begin{pNiceMatrix}[r,margin,first-row,first-col]
  & 0 & 1 & 2 & 3 & 4 & 5 & 6 & 7 \\
0 & 0 & 0 & 0 & 1 & 1 & 0 & 1 & 1 \\
5 & 0 & 0 & 0 & 0 & 0 & 1 & 0 & 0 \\
7 & 1 & 1 & 0 & 1 & 1 & 0 & 1 & 1 \\
3 & 1 & 1 & 0 & 1 & 0 & 0 & 0 & 0 \\
\end{pNiceMatrix}
\]
\caption{Generation of sub-matrix from full parity matrix}\label{fig:submatrixgen}
\end{figure}
To illustrate how CombSynth works, we'll work through an example, showing each of the steps explicitly, similar to what we did for RowCol. In doing so, we will look at the circuit in \eqref{quantumcircuithighlighted}, which has a quantum comb shown in \eqref{quantumcombunfolded}. We will apply this process without topological constraints as these aren't necessary for showing how our generalisation works. The quantum comb is made of 8 temporal qubits, however, our initial circuit is made of 4 logical ones. This means that only 4 of the 8 temporal qubits in our quantum comb can exist at any one time, and a $4\times8$ rectangular matrix will then be used for the elimination steps. The sub-matrix will initially take the form in Figure~\Ref{fig:submatrixgen}. Note that the rows are not in the same order as they were in the larger matrix: this is because they are placed in the position of their corresponding logical qubit. A table of the elimination steps is presented in Figure~\Ref{fig:submatrixrowop}, which highlights the row and column that get reduced to unit vectors and lists the row operations needed to do this. The elimination order of the temporal qubits is $5,7,6,3,4,0,1$.

\begin{figure}
  \begin{center}
    \begin{tabular}{|c|c|c|}
      \hline
      Initial sub-matrix & Eliminated sub-matrix & \makecell{Row operations \\ required for reduction}\\
      \hline
      $\submatrixAi$ & $\submatrixAii$ & N/A \\
      \hline
      $\submatrixBi$ & $\submatrixBii$ & \makecell{$R(7,0),R(3,7),$ \\ $R(4,7),R(0,7)$} \\
      \hline
      $\submatrixCi$ & $\submatrixCii$ & $R(6,4),R(3,6)$ \\
      \hline
      $\submatrixDi$ & $\submatrixDii$ & $R(3,4),R(0,3)$ \\
      \hline
      $\submatrixEi$ & $\submatrixEii$ & $R(0,4)$ \\
      \hline
      $\submatrixFi$ & $\submatrixFii$ & $R(0,2),R(1,0)$ \\
      \hline
      $\submatrixGi$ & $\submatrixGii$ & $R(1,2)$ \\
      \hline
    \end{tabular}
  \end{center}
  \caption{Row operations on sub matrices that reduce the parity matrix to the identity}\label{fig:submatrixrowop}
\end{figure}

\begin{figure}
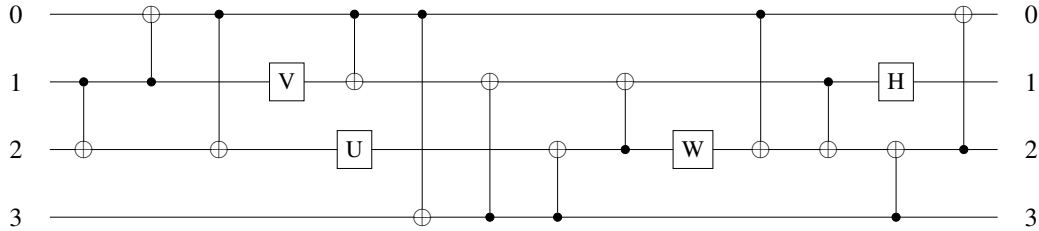

  \ctikzfig{generatedcircuit}
\caption{Circuit generated from the CombSynth procedure in Section~\Ref{sec:combsynth}}\label{fig:generatedcircuit}
\end{figure}

To complete the CombSynth algorithm, the comb generated from the row operations in Figure~\Ref{fig:submatrixrowop} is filled with the gates removed from the original circuit. This circuit, shown in Figure~\Ref{fig:generatedcircuit}, is smaller than the original, though only by 1 gate, but this effect grows as the circuit size increases due to more CNOT cancellations.

\section{Results}\label{sec:resandcon}
We have conducted a series of computational experiments to compare our generalisation of RowCol to a circuit slicing procedure. For the slicing we have used a na\"ive algorithm that cuts the circuit where the gates are found, similar to that shown in Figure~\Ref{slicecircB}. 20 random circuits with CNOT counts of $4,8,16,32,64,128,256,512$ and $1024$ were generated, and a set of non-CNOT gates distributed uniformly throughout them. The number of non-CNOT gates was set to be proportional to the number of CNOTs, with the proportionality factor varying from $5\%$ to $50\%$. The percentages of non-CNOTs in the table and graphs below are therefore the proportion of non-CNOTs to CNOTs, not the proportion of non-CNOTs to the total number of gates in the circuit. A range of architectures, popular for conducting similar computational experiments, was then selected to route our circuits onto. These were the 9-qubit square grid, 16-qubit square grid, Rigetti 16-qubit Aspen, 16-qubit IBM QX5 and 20-qubit IBM Tokyo. For each set of experimental parameters we recorded the proportional change in CNOT gates (CNOT overhead) due to the synthesis algorithms.

\graphicspath{{./figures}}
\begin{figure}[ht]
  \begin{subfigure}{0.33\textwidth}
    \includegraphics[scale=0.36]{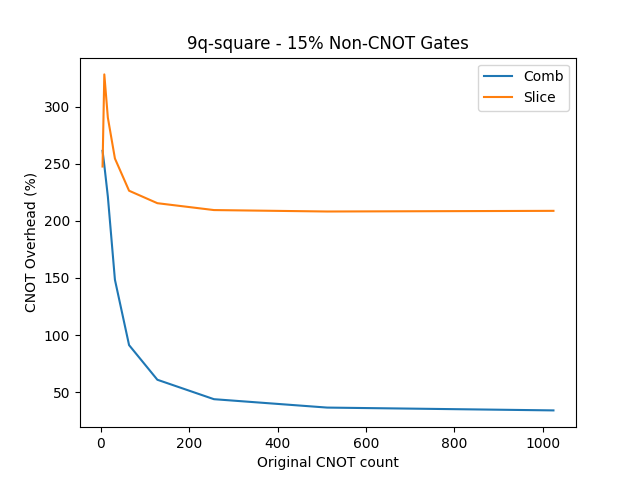}
    \caption{}\label{fig:9q-square-15}
  \end{subfigure}
  \begin{subfigure}{0.33\textwidth}
    \includegraphics[scale=0.36]{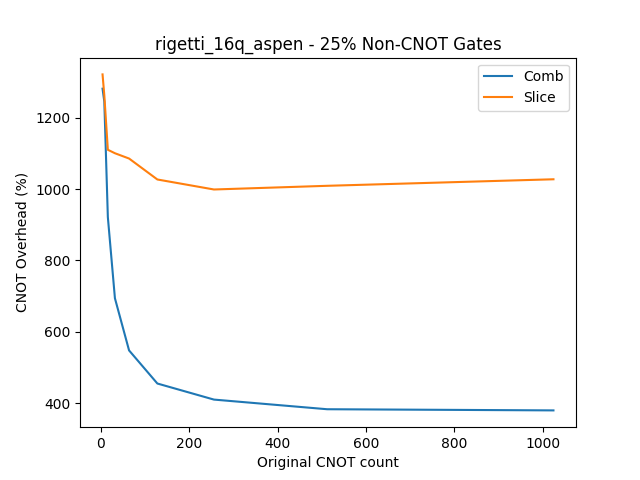}
    \caption{}\label{fig:rigetti_16q_aspen-25}
  \end{subfigure}
  \begin{subfigure}{0.33\textwidth}
    \includegraphics[scale=0.36]{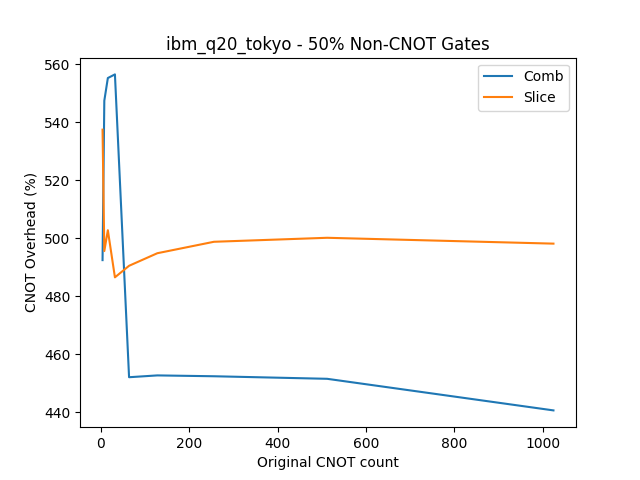}
    \caption{}\label{fig:ibm_q20_tokyo-50}
  \end{subfigure}
  \caption{Graphs showing the change in CNOT overhead when increasing the size of the circuit. This selection of graphs was chosen because it covers a wide range of the parameters of the computational experiment, however all the graphs still broadly have the same behaviour. They start off with a larger overhead, fluctuate a bit for the small circuit sizes, then approach some some constant value. With this approached value being smaller for the comb routing process then the slicing one. }\label{fig:realarchoverhead}
\end{figure}

A set of graphs with different parameters for the computational experiment is shown in Figure~\Ref{fig:realarchoverhead}. These graphs are illustrative of the behaviour of all of the experiments: the overhead is high for small circuits with the slicing procedure sometimes being better, but as the circuit size increases both overheads approach a constant value, with the comb algorithm outperforming the slicing one. This is to be expected as using quantum combs allows for better cancellations than the slicing process, due to each slice being routed independently. The overheads for the largest circuit sizes, 1024 CNOTs, are shown in Table~\Ref{tab:overhead}, giving a comparison of the asymptotic behaviour of each algorithm. It can be seen that larger proportions of non-CNOT gates reduce the advantage of using the comb algorithm: however it still outperforms the slicing method for all architectures and gate proportions tested.

\begin{table}[ht]
  \begin{center}
    \begin{NiceTabular}{*{9}{c}}[hvlines]
      \Block{2-1}{Architectures}       & \Block{1-2}{5\% \\ Non-CNOT Gates}  && \Block{1-2}{15\% \\ Non-CNOT Gates} && \Block{1-2}{25\% \\ Non-CNOT Gates} && \Block{1-2}{50\% \\ Non-CNOT Gates} & \\
                          & Comb & Slice & Comb & Slice & Comb & Slice & Comb & Slice \\
      9q-square           &-43.1\% &80.79\% &34.12\% &208.7\% &91.93\% &255.9\% &182.1\% &306.8\% \\
      16q-square          &13.11\% &344.5\% &154.4\% &511.1\% &263.9\% &564.6\% &437.0\% &606.9\% \\
\makecell{regetti_\\16q\_aspen} &47.31\% &555.4\% &231.2\% &893.1\% &379.6\% &1027\%  &614.9\% &1119\%  \\
      bm\_qx5             &32.27\% &461.9\% &197.2\% &698.8\% &322.4\% &783.6\% &527.8\% &837.7\% \\
      ibm\_q20\_tokyo     &33.17\% &393.1\% &183.6\% &481.0\% &289.3\% &500.2\% &440.7\% &498.2\% \\
    \end{NiceTabular}
  \end{center}
  \caption{CNOT overhead when routing to different architectures and proportions of non-CNOT gates. The values above are for the largest circuits tested: 1024 CNOTs}\label{tab:overhead}
\end{table}

\section{Conclusion and Future Work}\label{sec:futurework}
We have proposed an alternative to slicing the circuit, using quantum combs, for generalising synthesis algorithms to arbitrary circuits. This idea was then concretely outlined for the case of CNOT synthesis by developing a generalisation of RowCol that works for quantum combs, and showing this allows the routing of arbitrary circuits. Finally, through a series of computational experiments, we demonstrated that for large circuits our quantum comb generalisation of RowCol outperforms the slicing procedure on a range of architectures and  CNOT/non-CNOT proportions.
Work has recently been done on improving the performance of RowCol by allowing the qubits to be permuted by the synthesis algorithm~\cite{permrowcol}, investigating whether the performance of CombSynth could be improved in a similar way would be an interesting research direction. Currently, comb synthesis is designed to work with a subroutine that only works with one qubit (and hence one row/column) at a time, but it would be worth adapting it to work with synthesis algorithms which operate one more than one row or column at once, like the Patel-Markov-Hayes algorithm. Along a similar vein, it appears that our algorithm could be generalised to deal with multi-qubit holes by introducing some extra requirements than certain sets of temporal qubits should be extracted simultaneously.

Throughout this paper, we compared our quantum comb approach to that of slicing for generalising synthesis algorithms, however there have been some recent approaches that don't use slicing to perform circuit optimisation. These are lazy synthesis~\cite{martiel2022architectureaware} and ZX circuit extraction~\cite{duncan2020graph}, our approach differs from these as we focused on the utility of the higher order structure quantum combs for quantum compilation. Although a comparative analysis between our approach and these would be an interesting topic for further research. We focused on using quantum combs to generalise a CNOT synthesis algorithm in this paper: a natural next step would be to try and apply this idea to other synthesis algorithms such as the synthesis of Clifford circuits from stabiliser tableaux %~\cite{aaronson2004improved,maslov2018shorter,dehaene2003clifford}
or CNOT+phase circuits from phase polynomials. Finally, quantum combs allow for routing of circuits with ``black box'' operations, where you may not know what operation will eventually be performed at the time of compilation. This may be useful for compilation in the context of parameterised circuits such as those used in variational algorithms, or for classical-quantum algorithms where you want to route around mid-circuit measurements.

\bibliographystyle{eptcs}
\bibliography{references}

\end{document}